\documentclass[12pt]{article}
\usepackage{graphicx}

\def\hybrid{\topmargin 0pt      \oddsidemargin 0pt
        \headheight 0pt \headsep 0pt
       \voffset-1cm
        \textwidth 6.25in       
       \textheight 9.5in       
        \marginparwidth 0.0in
        \parskip 5pt plus 1pt   \jot = 1.5ex}
\catcode`\@=11
\def\marginnote#1{}

\newcount\hour
\newcount\minute
\newtoks\amorpm
\hour=\time\divide\hour by60
\minute=\time{\multiply\hour by60 \global\advance\minute by-\hour}
\edef\standardtime{{\ifnum\hour<12 \global\amorpm={am}%
        \else\global\amorpm={pm}\advance\hour by-12 \fi
        \ifnum\hour=0 \hour=12 \fi
        \number\hour:\ifnum\minute<10 0\fi\number\minute\the\amorpm}}
\edef\militarytime{\number\hour:\ifnum\minute<10 0\fi\number\minute}

\def\draftlabel#1{{\@bsphack\if@filesw {\let\thepage\relax
   \xdef\@gtempa{\write\@auxout{\string
      \newlabel{#1}{{\@currentlabel}{\thepage}}}}}\@gtempa
   \if@nobreak \ifvmode\nobreak\fi\fi\fi\@esphack}
        \gdef\@eqnlabel{#1}}
\def\@eqnlabel{}
\def\@vacuum{}
\def\draftmarginnote#1{\marginpar{\raggedright\scriptsize\tt#1}}

\def\draftlabel#1{{\@bsphack\if@filesw {\let\thepage\relax
   \xdef\@gtempa{\write\@auxout{\string
      \newlabel{#1}{{\@currentlabel}{\thepage}}}}}\@gtempa
   \if@nobreak \ifvmode\nobreak\fi\fi\fi\@esphack}
        \gdef\@eqnlabel{#1}}
\def\@eqnlabel{}
\def\@vacuum{}
\def\draftmarginnote#1{\marginpar{\raggedright\scriptsize\tt#1}}

\def\draft{\oddsidemargin -.5truein
        \def\@oddfoot{\sl preliminary draft \hfil
        \rm\thepage\hfil\sl\today\quad\militarytime}
        \let\@evenfoot\@oddfoot \overfullrule 3pt
        \let\label=\draftlabel
        \let\marginnote=\draftmarginnote
   \def\@eqnnum{(\theequation)\rlap{\kern\marginparsep\tt\@eqnlabel}%
\global\let\@eqnlabel\@vacuum}  }

\def\numberbysection{\@addtoreset{equation}{section}
        \def\theequation{\thesection.\arabic{equation}}}

\def\underline#1{\relax\ifmmode\@@underline#1\else
        $\@@underline{\hbox{#1}}$\relax\fi}

\def\titlepage{\@restonecolfalse\if@twocolumn\@restonecoltrue\onecolumn
     \else \newpage \fi \thispagestyle{empty}\c@page\z@
        \def\thefootnote{\fnsymbol{footnote}} }

\def\endtitlepage{\if@restonecol\twocolumn \else  \fi
        \def\thefootnote{\arabic{footnote}}
        \setcounter{footnote}{0}}  
\relax

\numberbysection
\hybrid

\newfont{\Bbb}{msbm10 scaled 1\@ptsize00}
\newfont{\Bbbb}{msbm7 scaled 1\@ptsize00}
\newcommand{\CC}{\mbox{\Bbb C}}

\newcommand{\DDD}{\raise-1pt\hbox{$\mbox{\Bbbb D}$}}

\newcommand{\UUU}{\raise-1pt\hbox{$\mbox{\Bbbb U}$}}

\newcommand{\ZZ}{\mbox{\Bbb Z}}
\newcommand{\z}{\raise-1pt\hbox{$\mbox{\Bbbb Z}$}}

\def\beq{\begin{equation}}
\def\eeq{\end{equation}}

\def\res{\mathop{\hbox{res}}\limits}

\newtheorem{theorem}{Theorem}[section]

\newtheorem{lemma-definition}{Lemma-Definition}[section]

\newtheorem{proposition}{Proposition}[section]

\begin{document}

\begin{titlepage}

\title{Quasi-periodic solutions of the universal hierarchy}

\author{\fbox{$\, \vphantom{A_A^A}$I. Krichever\thanks{
Columbia University, New York, USA}$\,\,$}
\and
A.~Zabrodin\thanks{
Skolkovo Institute of Science and Technology, 143026, Moscow, Russia and
National Research University Higher School of Economics,
20 Myasnitskaya Ulitsa,
Moscow 101000, Russia and
NRC ``Kurchatov institute'', Moscow, Russia;
e-mail: zabrodin@itep.ru}}

\date{August 2023}
\maketitle

\vspace{-7cm} \centerline{ \hfill ITEP-TH-19/23}\vspace{7cm}

\begin{abstract}

We construct quasi-periodic solutions of the universal 
hierarchy which includes the multi-component KP and Toda 
hierarchies and show how they fit into the bilinear formalism.
The tau-function is expressed in terms of the Riemann theta-function
multiplied by exponential function of 
a quadratic form in the hierarchical times.

\end{abstract}

\end{titlepage}


%

\tableofcontents


\section{Introduction}

Exact quasi-periodic solutions to nonlinear 
partial differential integrable equations (soliton equations) 
are known 
to be expressed in terms of 
Riemann's theta-functions associated to algebraic curves (Riemann surfaces). 
For the first time the explicit
formulas were obtained for the $(1+1)$-dimensional 
Korteveg-de Vries (KdV) equation in \cite{IM75,DMN76}. 
The unified approach to integration of more general $(2+1)$-dimensional Kadomtsev-Petviashvili
(KP) equation was developed
in \cite{Krichever77,Krichever77a} 
(see also the review \cite{Dubrovin81}). 
It was shown that the quasi-periodic
solutions are constructed starting from certain algebraic-geometrical data:
a smooth algebraic curve of finite genus 
with a marked point and a local parameter in the vicinity of this 
point. 

The methods developed in \cite{Krichever77,Krichever77a} 
are applicable to other integrable
equations as well. For example, quasi-periodic solutions to 
the 2D Toda lattice equation were first
constructed in \cite{Krichever81} starting from 
a smooth algebraic curve with two marked points. 

The basic object in the construction 
of the quasi-periodic (al\-geb\-ra\-ic-ge\-omet\-ri\-cal) solutions is the
Baker-Akhiezer function which can be regarded 
as a generalization of the exponential
function on the Riemann sphere to surfaces of nonzero genus. 
The key observation is that 
the Baker-Akhiezer function satisfies
an over-determined system of linear equations whose compatibility condition is equivalent
to the nonlinear equation. The Baker-Akhiezer function 
can be written explicitly
in terms of the Riemann theta functions and Abel map.

In the subsequent works of the Kyoto 
school \cite{DJKM83,JimboMiwa} the emphasis was made on the fact that the 
integrable equations form infinite hierarchies with infinitely many commuting flows. 
A common solution to the whole hierarchy is provided by the tau-function 
which depends on all independent variables (``times'') and satisfies
an infinite set of bilinear relations which can be encoded in one 
generating integral bilinear equation. 
The construction of algebraic-geometrical solutions can be 
extended to the hierarchy, the tau-function being the theta-function
multiplied by exponential function of a quadratic form in the times. 

The aim of this paper is to construct quasi-periodic solutions to the
hierarchy which we call universal because it contains both KP and 
Toda hierarchies and their multi-component extensions 
\cite{DJKM81}--\cite{UT84}. 
We show how the theta-functional solutions 
are embedded into the framework of the Kyoto school approach.
It turns out that equivalence of the two approaches 
is based on some nontrivial
facts about differentials on Riemann surfaces. 

The organization of the paper is as follows. Section 2 is devoted to
the bilinear approach to the universal hierarchy. In section 3,
following \cite{Springer,Fay,Mumford}, we present
basic facts from the theory of Riemann surfaces which are necessary for the
construction of the quasi-periodic solutions. In section 4 we introduce
the vector Baker-Akhiezer functions and give explicit formulas for them.
The core of the paper is section 5, where we prove that the 
algebric-geometrical tau-function (essentially the Riemann theta-function)
solves the integral bilinear equation of the universal 
hierarchy given in section 2. Section 6 is devoted to specialization
of this result to the multi-component KP and Toda hierarchies. 
Section 7 contains concluding remarks.

\section{The universal hierarchy}

The universal hierarchy is the multi-component KP hierarchy extended
by certain additional integrable flows. 
We call it universal because it contains
both KP and Toda hierarchies as well 
as their multi-component extensions.   

The independent variables are $n$ infinite sets of (in general complex) 
``times''
$$
{\bf t}=\{{\bf t}_1, {\bf t}_2, \ldots , {\bf t}_n\}, \qquad
{\bf t}_{\alpha}=\{t_{\alpha , 1}, t_{\alpha , 2}, t_{\alpha , 3}, \ldots \, \},
\qquad \alpha = 1, \ldots , n
$$
and $n$ additional variables $s_1, \ldots , s_n$
such that 
\beq\label{s1}
\sum_{\alpha =1}^n s_{\alpha}=0.
\eeq
We denote
${\bf s}=\{s_1, \ldots , s_n\}$
and
\beq\label{s2}
{\bf s}+[1]_{\alpha \beta} =\{s_1, \ldots , s_{\alpha}\! +\! 1, \ldots
s_{\beta}\! -\! 1, \ldots , s_n\}, \quad \alpha \neq \beta
\eeq
(i.e., $s_{\alpha}$ is changed to $s_{\alpha}+1$, $s_{\beta}$ is changed
to $s_{\beta}-1$ and all other variables remain unchanged).
For $\beta =\alpha$ we set ${\bf s}+[1]_{\alpha \alpha}={\bf s}$. 
We will also use the following standard
notation:
\beq\label{st1}
\left ({\bf t}\pm [k^{-1}]_{\gamma}\right )_{\alpha j}=t_{\alpha , j}\pm
\delta_{\alpha \gamma} \frac{k^{-j}}{j},
\eeq
\beq\label{st2}
\xi ({\bf t}_{\alpha}, k)=\sum_{j\geq 1}t_{\alpha , j}k^j.
\eeq
In general we treat $s_1, \ldots , s_n$ as complex variables.
If they are restricted to be integers, the hierarchy coincides with the 
one considered in
\cite{DJKM81}--\cite{Teo11}.

In the bilinear formalism, 
the dependent variable is the tau-function $\tau ({\bf s}, {\bf t})$ 
The $n$-component universal hierarchy is the infinite set of bilinear equations
for the tau-function which are encoded in the basic bilinear
relation
\beq\label{s3}
\begin{array}{l}
\displaystyle{
\sum_{\gamma =1}^n \epsilon_{\alpha \gamma}({\bf s})
\epsilon^{-1}_{\beta \gamma}({\bf s}')
\oint_{C_{\infty}}\! dk \, 
k^{s_{\gamma}-s_{\gamma}'+\delta_{\alpha \gamma}+\delta_{\beta \gamma}-2}
e^{\xi ({\bf t}_{\gamma}-{\bf t}_{\gamma}', \, k)}}
\\ \\
\displaystyle{\phantom{aaaaaaaaaaaaaaaaa}
\times \tau \left ({\bf s}+[1]_{\alpha \gamma}, 
{\bf t}-[k^{-1}]_{\gamma}\right )
\tau \left ({\bf s}'+[1]_{\gamma \beta}, {\bf t}'+[k^{-1}]_{\gamma}\right )=0}
\end{array}
\eeq 
valid for any ${\bf t}$, ${\bf t}'$, ${\bf s}$, ${\bf s}'$
such that ${\bf s}-{\bf s}' \in \ZZ ^n$
(and subject to the constraint (\ref{s1})). 
In (\ref{s3}) 
\beq\label{s3a}
\epsilon_{\alpha \gamma}({\bf s})=\left \{
\begin{array}{cl} 
\displaystyle{\exp \, \Bigl (-i\pi \!\! \sum_{\alpha <\mu \leq \gamma}
\!\! s_{\mu}
\Bigr )}, 
&\quad \alpha < \gamma
\\ 
\hspace{-1.5cm}1, &\quad \alpha =\gamma
\\ 
\displaystyle{-\, \vphantom{\sum^{\alpha \leq }}
\exp \, \Bigl (i\pi \!\! \sum_{\gamma <\mu \leq \alpha}\!\! s_{\mu}
\Bigr )}, 
&\quad \alpha > \gamma .
\end{array}
\right.
\eeq
The contour $C_{\infty}$ is a big circle around infinity. 
Different bilinear relations for the tau-function which follow from
(\ref{s3}) for special choices of ${\bf s}-{\bf s}'$ and ${\bf t}-{\bf t}'$
are given in \cite{Teo11}. 

Let us introduce the wave function and its dual:
\beq\label{s4}
\begin{array}{l}
\displaystyle{\psi_{\alpha \beta}({\bf s}, {\bf t},k)=
\epsilon_{\alpha \beta}({\bf s})\,
\frac{\tau \left ({\bf s}+[1]_{\alpha \beta},
{\bf t}-[k^{-1}]_{\beta}\right )}{\tau ({\bf s}, {\bf t})}\,
k^{s_{\beta}+\delta_{\alpha \beta}-1}e^{\xi ({\bf t}_{\beta}, k)},
}
\\ \\
\displaystyle{\psi^{*}_{\alpha \beta}({\bf s}, {\bf t},k)=
\epsilon_{\beta \alpha}^{-1}({\bf s})\,
\frac{\tau \left ({\bf s}+[1]_{\alpha \beta},
{\bf t}+[k^{-1}]_{\alpha}\right )}{\tau ({\bf s}, {\bf t})}\,
k^{-s_{\alpha}+\delta_{\alpha \beta}-1}e^{-\xi ({\bf t}_{\alpha}, k)}.
}
\end{array}
\eeq
In the right hand sides, the factor $k^{s_{\beta}}$ should be 
understood as $k^{s_{\beta}}=e^{s_{\beta}\log k}$ with a cut from
$\infty$ to $0$ in the $k$-plane, and the wave functions 
as functions of $k$ have jumps
across the cut. 

Note that the bilinear relation for the tau-function can be written 
as the following bilinear relation for the wave functions:
\beq\label{s5}
\sum_{\gamma}\oint_{C_{\infty}}\! 
\psi_{\alpha \gamma} ({\bf s}, 
{\bf t},k)\psi^{*}_{\gamma \beta} ({\bf s}', {\bf t}',k)dk =0,
\quad {\bf s}-{\bf s}' \in \ZZ ^n.
\eeq
Note that if ${\bf s}-{\bf s}' \in \ZZ ^n$, the product 
$\psi_{\alpha \gamma} ({\bf s}, 
{\bf t},k)\psi^{*}_{\gamma \beta} ({\bf s}', {\bf t}',k)$ has no jump
and the integral is well-defined.
Introducing $n\! \times \! n$ matrices $\hat \psi$, $\hat \psi^*$
with matrix elements $\psi_{\alpha \beta}$, $\psi^*_{\alpha \beta}$,
we can rewrite (\ref{s5}) in the matrix form:
\beq\label{s6}
\oint_{C_{\infty}}\hat \psi ({\bf s}, {\bf t}, k)
\hat \psi^* ({\bf s}', {\bf t}', k)dk=0,
\quad {\bf s}-{\bf s}' \in \ZZ ^n.
\eeq

In this paper we construct algebraic-geometrical (quasi-periodic) 
solutions to the bilinear
relation (\ref{s3}).

\section{The necessary facts from the theory of Riemann surfaces}

Let $\Gamma$ be a smooth compact algebraic curve (a Riemann surface) 
of genus $g$. We fix a canonical basis of
cycles ${\sf a}_{\alpha}, {\sf b}_{\alpha}$ ($\alpha =1, \ldots , g$) with the intersections
${\sf a}_{\alpha}\circ {\sf a}_{\beta}={\sf b}_{\alpha}\circ {\sf b}_{\beta}=0$,
${\sf a}_{\alpha}\circ {\sf b}_{\beta}=\delta_{\alpha \beta}$.
We will sometimes denote the sets $\{{\sf a}_{\alpha}\}$, 
$\{{\sf b}_{\alpha}\}$ as ``vectors'' $\vec {\sf a}$, $\vec {\sf b}$
with $g$ ``components''. 
We also fix a basis of holomorphic 
differentials $d\omega_{\alpha}$  
normalized by the condition
$\displaystyle{\oint_{{\sf a}_{\alpha}}
d\omega_{\beta}=\delta_{\alpha \beta}}$. 
The period matrix is defined as
\beq\label{qp1}
T_{\alpha \beta}=\oint_{{\sf b}_{\alpha}}d\omega_{\beta}, \qquad \alpha , \beta =1, \ldots , g.
\eeq
It is a symmetric matrix with positively defined imaginary part.
The Jacobian of the curve $\Gamma$ is the $g$-dimensional complex torus
\beq\label{qp4}
J(\Gamma )=\CC ^g /\{\vec N +T \vec M\},
\eeq
where $\vec N$, $\vec M$ are $g$-dimensional vectors with integer 
components. 

\paragraph{The Riemann theta-functions.}
The Riemann theta-function associated with the Riemann 
surface is defined by the 
absolutely convergent series
\beq\label{qp2}
\Theta(\vec z)=\Theta(\vec z|T)=
\sum_{\vec n \in \z ^{g}}e^{\pi i (\vec n, T\vec n)+2\pi i (\vec n, \vec z)},
\eeq
where $\vec z=(z_1, \ldots , z_g)$ and $\displaystyle{(\vec n, \vec z)=
\sum_{\alpha =1}^g n_{\alpha}z_{\alpha}}$.
It is an entire function with the following quasi-periodicity property:
\beq\label{qp2a}
\Theta (\vec z +\vec N +T\vec M )=\exp (-\pi i (\vec M, T\vec M)-
2\pi i (\vec M, \vec z)) \Theta (\vec z).
\eeq
More generally, one can introduce the theta-functions with characteristics
$\vec \delta'$, $\vec \delta'$:
\beq\label{char}
\Theta \left [\begin{array}{c}\vec \delta ' \\ \vec \delta^{''}
\end{array}\right ] (\vec z)
=\sum_{\vec n \in \z ^{g}}e^{\pi i (\vec n +\vec \delta ', 
T(\vec n +\vec \delta '))+2\pi i (\vec n+\vec \delta ', 
\vec z+\vec \delta '')}.
\eeq

\paragraph{The Abel map.}
Fix a point $Q_0\in \Gamma$ and define the Abel map $\vec A(P)$, $P\in \Gamma$
from $\Gamma$ to $J(\Gamma )$ as
\beq\label{qp3}
\vec A(P)=
\int_{Q_0}^P d \vec \omega , \qquad d\vec 
\omega =(d\omega_1, \ldots , d\omega_g ).
\eeq
The Abel map can be extended to the group of divisors ${\cal D}=n_1Q_1+\ldots +n_KQ_K$ as
\beq\label{qp3a}
\vec A({\cal D})=\sum_{i=1}^K n_i\int_{Q_0}^{Q_i} d \vec \omega =\sum_{i=1}^K
n_i\vec A(Q_i).
\eeq

\paragraph{Riemann's constants.}
Consider the function $f(P)=\Theta (\vec A(P)-\vec e)$ and assume that it is not 
identically zero. It can be shown that this function has $g$ zeros on $\Gamma$
at a divisor ${\cal D}=Q_1+ \ldots +Q_g$ and $\vec A( {\cal D})=\vec e -\vec K$,
where $\vec K=
(K_1, \ldots , K_g)$ is the 
vector of Riemann's constants
\beq\label{qp7}
K_{\alpha}= \pi i +\pi i T_{\alpha \alpha}-2\pi i
\sum_{\beta \neq \alpha}\oint_{a_{\beta}}\omega_{\alpha} (P)d\omega_{\beta}(P).
\eeq
In other words,
for any non-special effective divisor
${\cal D}=Q_1+\ldots +Q_g$ of degree $g$ the function
$$
f(P)=\Theta \Bigl (\vec A(P)-\vec A({\cal D}) -\vec K\Bigr )
$$
has exactly $g$ zeros at the points $Q_1, \ldots , Q_g$.  
Let ${\cal K}$ be the canonical class of divisors (the equivalence class of divisors
of poles and zeros of abelian differentials on $\Gamma$), then one can show that
\beq\label{qp8}
2\vec K=-\vec A({\cal K}).
\eeq
It is known that $\mbox{deg}\, {\cal K}=2g-2$. In particular, this means that 
holomorphic differentials have $2g-2$ zeros on $\Gamma$.

\paragraph{Abelian differentials of the second kind.}
Let $P_{\alpha}\in \Gamma$, $\alpha =1, \ldots , n$ be 
marked points (punctures) and $k_{\alpha}^{-1}$ be local parameters 
in neighborhoods of the marked points ($k_{\alpha}=\infty$ at $P_{\alpha}$). 
Let $d\Omega_j$ 
be differentials of the second kind with the only pole at $P_{\alpha}$ of the form
$$
d\Omega_j^{(\alpha )} = dk_{\alpha}^j +O(k_{\alpha}^{-2})dk_{\alpha}, 
\quad k_{\alpha}\to \infty
$$
normalized by the condition $\displaystyle{\oint_{\vec {\sf a}}
d\Omega_{j}^{(\alpha )}=0}$, and $\Omega_j^{(\alpha )}$
be the (multi-valued) functions
$$
\Omega_j^{(\alpha )}(P)=\int_{Q_0}^{P}d\Omega_j^{(\alpha )} +q^{(\alpha )}_j,
$$
where the constants $q_j^{(\alpha )}$ are chosen in such a 
way that $\Omega_i^{(\alpha )}(P)=k_{\alpha}^i +O(k_{\alpha}^{-1})$, namely,
\beq\label{qp3b}
\Omega_i^{(\alpha )}(P)=k_{\alpha}^i +
\sum_{j\geq 1} \frac{1}{j}\, \Omega_{ij}^{(\alpha \alpha)}k_{\alpha}^{-j}.
\eeq
In the vicinity of another marked point, $P_{\beta}$, we have:
\beq\label{qp3c}
\Omega_i^{(\alpha )}(P)=\Omega_i^{(\alpha )}(P_{\beta})+
\sum_{j\geq 1} \frac{1}{j}\, \Omega_{ij}^{(\alpha \beta)}k_{\beta}^{-j}.
\eeq
It follows from the Riemann bilnear identity that 
\beq\label{qp3d}
\Omega_{ij}^{(\alpha \beta )}=\Omega_{ji}^{(\beta \alpha )}. 
\eeq
Set
\beq\label{qp5}
\vec U_j^{(\alpha )}=\frac{1}{2\pi i}\oint_{\vec {\sf b}}
d\Omega_j^{(\alpha )}.
\eeq
One can prove that
\beq\label{qp6}
\vec A(P)-\vec A(P_{\alpha })=
-\sum_{j\geq 1}\frac{1}{j}\, \vec U_j^{(\alpha )} k_{\alpha}^{-j}.
\eeq

\paragraph{Abelian differentials of the third kind.}
We will also need abelian differentials of the third kind 
$d\Omega_0^{(\alpha \beta )}$ which are meromorphic differentials
with two simple poles at the marked points 
$P_{\alpha}$, $P_{\beta}$ with residue $1$ 
at $P_{\alpha}$ and residue $-1$ at $P_{\beta}$ (dipole differentials). 
The condition that they are normalized, i.e. have zero ${\sf a}$-periods, 
fixes them uniquely. 
Clearly, $d\Omega_0^{(\alpha \beta )}=-d\Omega_0^{( \beta \alpha )}$.
Explicitly, the differential $d\Omega_0^{(\alpha \beta )}$ is given by
\beq\label{qp9a}
d\Omega_0^{(\alpha \beta )}=d\log 
\frac{\Theta_{*}(\vec A(P)-
\vec A(P_{\alpha}))}{\Theta_{*}(\vec A(P)-\vec A(P_{\beta}))},
\eeq
where $\Theta_{*}$ is the Riemann theta-function with an odd half-integer
characteristics. 
The expansions of the function 
$\Omega_0^{(\alpha \beta )}(P)$ near the points $P_{\alpha}$, $P_{\beta}$ 
are
\beq\label{qp9}
\Omega_0^{(\alpha \beta )}(P)=\left \{
\begin{array}{l}
\displaystyle{
-\log k_{\alpha} +\Omega_{00}^{(\alpha \beta )}+\sum_{j\geq 1}\frac{1}{j}\, 
\Omega_{0j}^{(\alpha \beta )}k_{\alpha}^{-j}, \quad P\to P_{\alpha},}
\\ \\
\displaystyle{
\log k_{\beta} -\Omega_{00}^{(\beta \alpha )}-\sum_{j\geq 1}\frac{1}{j}\, 
\Omega_{0j}^{(\beta \alpha )}k_{\beta}^{-j}, \quad P\to P_{\beta},}
\end{array}
\right.
\eeq
where
\beq\label{qp9b}
\Omega_{00}^{(\alpha \beta )}=\log C_{\alpha} -\log
\Theta_{*}(\vec A(P_{\beta})-\vec A(P_{\alpha})), \quad
C_{\alpha}=\sum_{\nu =1}^g \Theta_{*, \nu}(\vec 0)U_1^{(\alpha ), \nu}.
\eeq
Set
\beq\label{qp10}
\vec U_0^{(\alpha \beta )}=\frac{1}{2\pi i}\oint_{\vec 
{\sf b}}d\Omega_0^{(\alpha \beta )}.
\eeq
The following important relations are 
immediate consequences of the Riemann bilinear
identity:
\beq\label{qp11}
\vec A(P_\alpha )-\vec A(P_{\beta})=\vec U_0^{(\alpha \beta )},
\eeq
\beq\label{qp12}
\Omega^{(\alpha )}_i(P_\beta )=\Omega_{0i}^{(\alpha \beta )}.
\eeq
It is assumed that the integration contours in the left hand sides
do not intersect the ${\sf a}$- and ${\sf b}$-cycles.

\paragraph{The holomorphic differential $d\zeta$.}
In the same way as in \cite{Z23}, one can obtain the relation
\beq\label{stau7a}
\exp 
\Bigl (-\sum_{i,j} \Omega_{ij}^{(\gamma \gamma )}\,
\frac{k_{\gamma}^{-i-j}}{ij}\Bigr )dk_{\gamma}=
\frac{C_{\gamma}\, d\zeta (P)}{\Theta_{*}^2(\vec A(P)-\vec A(P_{\gamma}))},
\eeq
where $d\zeta$ is the holomorphic differential
\beq\label{qp86}
d\zeta =\sum_{\alpha =1}^g \Theta_{*, \alpha}(\vec 0)d\omega_{\alpha}.
\eeq
It is known (see \cite{Mumford}) that the differential 
$d\zeta$ has double zeros at some
$g-1$ points $R_1, \ldots , R_{g-1}$ while the function
$$
f_{*}(P)= \Theta_{*}\Bigl ( \vec A(P)\! -\! \vec A(P_{\gamma} )\Bigr )
$$
has simple zeros at the same points $R_i$ and $P_{\gamma}$. Therefore, the differential
in the right hand side of (\ref{stau7a}) has the only (second order) pole at $P_{\gamma}$
and no zeros. However, this differential is well-defined only on a covering of the curve
$\Gamma$ because it is not single-valued.

\section{The vector Baker-Akhiezer functions}

Let us fix $n$ marked points $P_1, \ldots , P_n$ (punctures) 
on the curve $\Gamma$ and make $n$ (non-intersecting) 
cuts ${\sf c}_{\alpha}$ from a point $Q_0$
(the initial point of the Abel transform) to the marked points $P_{\alpha}$.
We also fix $g+n-1$ points $Q_1 , \ldots , Q_{g+n-1}$ 
in general position. The corresponding effective divisor is
$$
{\cal D}=Q_1+Q_2+\ldots +Q_{g+n-1}.
$$

\begin{proposition}
There exists a unique
function $\Psi_{\alpha}({\bf s},{\bf t}, P)$ on $\Gamma$ such that:
\begin{itemize}
\item[$1^{0}$.] The function $\Psi_{\alpha}$ is meromorphic outside the
punctures and cuts and has at most simple poles at the points of the divisor
${\cal D}$ (if all of them are distinct);
\item[$2^{0}$.] The boundary values $\Psi_{\alpha}^{(\pm )}
({\bf s},{\bf t}, P)$ of the function $\Psi_{\alpha}$ 
on the different sides 
of the cut ${\sf c}_{\beta}$ satisfy the relation
\beq\label{v0}
\Psi_{\alpha}^{(+ )}
({\bf s},{\bf t}, P)=e^{2\pi i s_{\beta}}\Psi_{\alpha}^{(-)}
({\bf s},{\bf t}, P), \qquad P\in {\sf c}_{\beta};
\eeq
\item[$3^{0}$.] In the neighborhood of the puncture $P_{\beta}$ it has the 
form
\beq\label{v1}
\Psi_{\alpha}({\bf s},{\bf t}, P)=k_{\beta}^{s_{\beta}}
e^{\xi ({\bf t}_{\beta}, k_{\beta})}
\left (\delta_{\alpha \beta} +\sum_{j\geq 1}\xi_{\alpha \beta , j}
({\bf s}, {\bf t})k_{\beta}^{-j}\right ).
\eeq
\end{itemize}
\end{proposition}

\noindent
The proof of propositions of this type is standard in the theory
of Riemann surfaces.
The vector function $\Psi$ 
with components $\Psi_{\alpha}$ 
is called the vector Baker-Akhiezer function. 

We also need the definition of the dual Baker-Akhiezer function.
Let $${\cal D}^{*}=Q_1^{*}+ \ldots + Q_{g+n-1}^{*}$$ be the effective 
divisor satisfying the condition
\beq\label{v2}
{\cal D}+{\cal D}^{*}={\cal K}+2\sum_{\gamma =1}^n P_{\gamma},
\eeq
where ${\cal K}$ is the canonical class. 
In terms of the Abel map, this condition reads
\beq\label{v2a}
\vec A({\cal D})+\vec A({\cal D}^{*})+2\vec K=2\sum_{\gamma =1}^n 
\vec A(P_{\gamma}),
\eeq
where $\vec K$ is the vector of Riemann's constants. 

\begin{proposition}
There exists a unique function $\Psi_{\alpha}^{*}({\bf s},
{\bf t}, P)$ such that:
\begin{itemize}
\item[$1^{0}$.] The function $\Psi_{\alpha}^{*}$ is meromorphic outside the
punctures and cuts and has at most simple poles at the points of the divisor
${\cal D}^{*}$ (if all of them are distinct);
\item[$2^{0}$.] The boundary values $\Psi_{\alpha}^{*(\pm )}
({\bf s},{\bf t}, P)$ of the function $\Psi_{\alpha}^{*}$ 
on the different sides 
of the cut ${\sf c}_{\beta}$ satisfy the relation
\beq\label{v0a}
\Psi_{\alpha}^{*(+ )}
({\bf s},{\bf t}, P)=e^{-2\pi i s_{\beta}}\Psi_{\alpha}^{*(-)}
({\bf s},{\bf t}, P), \qquad P\in {\sf c}_{\beta};
\eeq
\item[$3^{0}$.] In the neighborhood of the puncture $P_{\beta}$ it has the 
form
\beq\label{v1a}
\Psi_{\alpha}^{*}({\bf s},{\bf t}, P)=k_{\beta}^{-s_{\beta}}
e^{-\xi ({\bf t}_{\beta}, k_{\beta})}
\left (\delta_{\alpha \beta} +\sum_{j\geq 1}\xi^{*}_{\alpha \beta , j}
({\bf s},{\bf t})k_{\beta}^{-j}\right ).
\eeq
\end{itemize}
\end{proposition}

\noindent
The vector function $\Psi^{*}$ 
with components $\Psi^{*}_{\alpha}$ 
is called the dual vector Baker-Akhiezer function.

In \cite{KBBT95} an explicit formula for the vector 
Baker-Akhiezer function in the case 
${\bf s}=0$ in terms of Riemann's theta-functions was given.
Here we extend it to the case of arbitrary ${\bf s}$. 

According to the Riemann-Roch theorem, for any effective divisor ${\cal D}$
of degree $g+n-1$ in general position there exists a 
unique meromorphic function $h_{\alpha}(P)$ such that 
the divisor of its poles coincides with ${\cal D}$ and such that
$h_{\alpha}(P_{\beta})=\delta_{\alpha \beta}$. Explicitly, this function 
can be written as follows:
\beq\label{v3}
h_{\alpha}(P)=\frac{f_{\alpha}(P)}{f_{\alpha}(P_{\alpha})},
\quad
f_{\alpha}(P)=\Theta (\vec A(P)-\vec A(P_{\alpha})+\vec Z)
\frac{\prod\limits_{\gamma \neq \alpha}
\Theta (\vec A(P)+\vec R_{\gamma})}{\prod\limits_{\nu =1}^{n}
\Theta (\vec A(P)+\vec S_{\nu})},
\eeq
where
\beq\label{v4}
\vec R_{\alpha}=-\vec K -\vec A(P_{\alpha})-\sum_{s=1}^{g-1}
\vec A(Q_s),
\eeq
\beq\label{v5}
\vec S_{\alpha}=-\vec K -\vec A(Q_{g-1+\alpha})-\sum_{s=1}^{g-1}
\vec A(Q_s),
\eeq
and 
\beq\label{v6}
\vec Z = -\vec K -\vec A({\cal D})+\sum_{\nu =1}^n \vec A(P_{\nu} ).
\eeq
Besides $n-1$ zeros at the points $P_{\gamma}$, $\gamma \neq \alpha$,
the function $h_{\alpha}(P)$ has other $g$ simple zeros at some
points $p_1^{(\alpha )}, \ldots , p_g^{(\alpha )}$ which are zeros of
$\Theta (\vec A(P) -\vec A(P_{\alpha })+\vec Z)$.
We also introduce the functions $h_{\alpha}^*(P)$ 
constructed in a similar way using the points of the divisor ${\cal D}^*$.
The function $h_{\alpha}^*(P)$ has $n-1$ zeros 
at the points $P_{\gamma}$, $\gamma \neq \alpha$ and
other $g$ simple zeros at some
points $p_1^{*(\alpha )}, \ldots , p_g^{*(\alpha )}$
which are zeros of
$\Theta (\vec A(P) -\vec A(P_{\alpha })+\vec Z^*)$. Note that
$\displaystyle{\vec Z^* = -\vec K -\vec A({\cal D}^*)+
\sum_{\nu =1}^n \vec A(P_{\nu})=-\vec Z}$.

In order to write down the explicit expressions for the vector 
Baker-Akhiezer function and its dual in a compact form, we prepare
some convenient notation. First, we denote
\beq\label{v6a}
F_{\alpha}(P, {\bf s}, {\bf t}, \vec Z)=
\frac{\Theta (\vec A(P)-\vec A(P_{\alpha}) +\vec U({\bf t}) 
-\vec A({\bf s}) +\vec Z)}{\Theta (\vec U({\bf t}) 
-\vec A({\bf s}) +\vec Z)},
\eeq
where
\beq\label{v6b}
\vec U({\bf t})=\sum_{\mu }\sum_{j\geq 1}\vec U_j^{(\mu )}t_{\mu , j},
\qquad
\vec A({\bf s})=\sum_{\nu =1}^n s_{\nu}\vec A(P_{\nu}).
\eeq
Second, we denote
\beq\label{v6c}
\hat \Omega_{\alpha}(P, {\bf t})=\sum_{\mu}\sum_j \Omega_j^{(\mu )}
(P) t_{\mu , j}-\sum_{\mu \neq \alpha}\sum_j \Omega_j^{(\mu )}
(P_{\alpha}) t_{\mu , j},
\eeq
\beq\label{v6d}
\hat \Omega_{\alpha}^{(0)}(P, {\bf s})=\sum_{\mu} 
\Omega_0^{(0\mu )}(P) s_{\mu} -\sum_{\mu \neq \alpha} 
\Omega_0^{(0\mu )}(P_{\alpha}) s_{\mu}+\Omega_{00}^{(\alpha 0)}s_{\alpha},
\eeq
where $d\Omega_0^{(0\mu )}$ is the normalized dipole differential with 
first order poles at the points $Q_0$ (the initial point of the Abel transform)
and $P_{\mu}$. The condition (\ref{s1}) implies that 
$\hat \Omega_{\alpha}^{(0)}(P, {\bf s})$ does not depend on $Q_0$. 
Using this notation, the vector Baker-Akhiezer function and its dual
can be written in the form
\beq\label{v7}
\Psi_{\alpha}({\bf s}, {\bf t}, P)=h_{\alpha}(P)
\frac{F_{\alpha}(P, {\bf s}, {\bf t}, \vec Z)}{F_{\alpha}(P, 0, 0, \vec Z)}
\, e^{\hat \Omega_{\alpha}(P, {\bf t})+
\hat \Omega_{\alpha}^{(0)}(P, {\bf s})},
\eeq
\beq\label{v8}
\Psi^{*}_{\beta}({\bf s}, {\bf t}, P)=h^{*}_{\beta}(P)
\frac{F_{\beta}(P, -{\bf s}, -{\bf t}, -\vec Z)}{F_{\beta}(P, 0, 0, -\vec Z)}
\, e^{-\hat \Omega_{\beta}(P, {\bf t})-
\hat \Omega_{\beta}^{(0)}(P, {\bf s})}.
\eeq

Let $d\Omega (P)$ be the meromorphic differential with second order poles
at the punctures $P_{\alpha}$ and zeros at the divisors ${\cal D}$, 
${\cal D}^*$. Such a differential exists because of the condition (\ref{v2}).
Moreover, it is unique up to multiplication by a constant factor. 
Indeed, the linear space of all differentials with second order poles
at the punctures has dimension $g+2n-1$ (there are 
$n$ independent coefficients 
in front of $dk_{\alpha}$, $\alpha =1, \ldots , n$, $n-1$ independent 
coefficients in front of $dk_{\alpha}/k_{\alpha}$ and $g$ coefficients
in front of the holomorphic differentials $d\omega_{\nu}$)
but fixing the positions of zeros 
imposes $g+2n-2$ conditions (the $2(g+n-1)$ zeros 
are subject to $g$ conditions (\ref{v2a})), 
so the space of such differentials with 
fixed zeros is one-dimensional. Alternatively, one may fix the divisor
${\cal D}$ and impose the condition that near each point 
$P_{\alpha}$ the differential $d\Omega$ behaves as 
$d\Omega (P)=dk_{\alpha}+\ldots$. These conditions fix the dual divisor
${\cal D}^*$. 

Let us consider the differential
\beq\label{eta1}
d\eta^{(\alpha \beta )} =
\frac{F_{\alpha}(P, 0, 0, \vec Z)
F_{\beta} (P, 0, 0, -\vec Z)\, 
d\zeta (P)}{\Theta_{*}(\vec A(P)-\vec A(P_{\alpha}))
\Theta_{*}(\vec A(P)-\vec A(P_{\beta}))}\, ,
\eeq
where $d\zeta (P)$ is the holomorphic differential
$$
d\zeta = \sum_{\alpha =1}^g \Theta_{*, \alpha}(0)d\omega_{\alpha}.
$$
If $\beta \neq \alpha$, then $d\eta^{(\alpha \beta )}$ is the 
well-defined dipole
differential with the only first order 
poles at $P_{\alpha}$, $P_{\beta}$ and $2g$ 
zeros at the points $p_1^{(\alpha )}, \ldots , p_g^{(\alpha )}$,
$p_1^{*(\beta )}, \ldots , p_g^{*(\beta )}$.
If $\beta =\alpha$,
then $d\eta^{(\alpha \alpha )}$ is the differential of the second kind with
the only second order pole at $P_{\alpha}$ and zeros at 
the points $p_1^{(\alpha )}, \ldots , p_g^{(\alpha )}$,
$p_1^{*(\alpha )}, \ldots , p_g^{*(\alpha )}$.
It is easy to see that 
the differential 
$$\frac{d\eta ^{(\alpha \beta )}(P)}{h_{\alpha}(P)h^{*}_{\beta}(P)}$$
has second order poles at all the marked points $P_{\gamma}$ 
and simple zeros at 
the divisors ${\cal D}$, 
${\cal D}^*$. Since the space of such differentials is one-dimensional
(see above), we conclude that 
\beq\label{eta2}
d\Omega (P)=c_{\alpha \beta}
\frac{d\eta ^{(\alpha \beta )}(P)}{h_{\alpha}(P)
h^{*}_{\beta}(P)},
\eeq
where $c_{\alpha \beta}$ is a numerical constant. 

Clearly, at ${\bf s}-{\bf s}'\in \ZZ ^n$
the differential $\Psi_{\alpha}({\bf s}, {\bf t}, P)
\Psi_{\beta}^{*}({\bf s}', {\bf t}', P)d\Omega (P)$ is well-defined on
$\Gamma$ and 
has singularities only at 
the points $P_{\gamma}$. The sum of its residues must be zero
for all ${\bf s}, {\bf s}'$, ${\bf t}, {\bf t}'$ (such that 
${\bf s}-{\bf s}'\in \ZZ ^n$):
\beq\label{v10}
\sum_{\gamma} \res_{P_{\gamma}}\, \Bigl (\Psi_{\alpha}({\bf s}, {\bf t}, P)
\Psi_{\beta}^{*}({\bf s}', {\bf t}', P)d\Omega (P)\Bigr )=0, 
\qquad
{\bf s}-{\bf s}'\in \ZZ ^n.
\eeq
We will show that for algebraic-geometrical solutions the bilinear
relation (\ref{s5}) for the wave functions is equivalent to (\ref{v10}).

\section{Tau-function of algebraic-geometrical solutions to the universal
hierarchy}

This section is devoted to the proof of the following theorem.

\begin{theorem}
The tau-function  
\beq\label{stau1}
\tau ({\bf s}, {\bf t})=
e^{-Q ({\bf s}, {\bf t})}\, \Theta \Bigl ( \vec U({\bf t})-
\vec A({\bf s})+\vec Z\Bigr )
\eeq
for algebraic-geometrical solutions 
to the universal hierarchy, where
$Q({\bf s}, {\bf t})$ is the quad\-ra\-tic form
\beq\label{stau2}
Q({\bf s}, {\bf t})=\frac{i\pi}{4}\sum_{\mu}
s_{\mu}^2+\frac{1}{2}\sum_{\mu \neq \nu}\Omega_{00}^{(\mu \nu )}
s_{\mu}s_{\nu}+\sum_{\mu \neq \nu}\sum_j \Omega_{0j}^{(\mu \nu )}
s_{\nu}t_{\mu , j}+
\frac{1}{2}\sum_{\mu , \nu}\sum_{i,j}
\Omega_{ij}^{(\mu \nu )}t_{\mu , i}t_{\nu , j},
\eeq
solves the bilinear relation
(\ref{s5}).
\end{theorem}

\noindent
The rest of this section contains the proof that the bilinear
relation (\ref{s3}) with $\tau$ as in (\ref{stau1}) is the same as 
the identity (\ref{v10}). 

Our strategy is to find the wave functions 
$\psi_{\alpha \gamma}$ and $\psi^*_{\gamma \beta}$
according to formulas (\ref{s4}) and compare with expansions of the
Baker-Akhiezer functions $\Psi_{\alpha}$, $\Psi^*_{\beta}$ near the
point $P_{\gamma}$. 
At $\alpha \neq \gamma$, a direct calculation yields
\beq\label{stau3}
\begin{array}{l}
\displaystyle{
\psi_{\alpha \gamma}({\bf s}, {\bf t}, k_{\gamma})=
\epsilon_{\alpha \gamma}({\bf s})
\exp \left (-\frac{i\pi}{2}(s_{\alpha}\! -\! s_{\gamma}\! +\! 1)+
\hat \Omega_{\alpha}(P, {\bf t})+
\sum_{\mu \neq \gamma}\Omega_0^{(\gamma \mu )}(P)s_{\mu}-
\Omega_0^{(\alpha \gamma )}(P) \right.}
\\ \\
\phantom{aaaaaaa}\displaystyle{
\left. -\frac{1}{2}\sum_{\mu \neq \alpha}(\Omega_{00}^{(\mu \alpha )}
+\Omega_{00}^{(\alpha \mu )})s_{\mu}+
\frac{1}{2}\sum_{\mu \neq \gamma}(\Omega_{00}^{(\mu \gamma )}
-\Omega_{00}^{(\gamma \mu )})s_{\mu}+\frac{1}{2}(\Omega_{00}^{(\alpha \gamma )}
-\Omega_{00}^{(\gamma \alpha )})\right )}
\\ \\
\phantom{aaaaaaaaaaaaaaaaaaaaaaaa}\displaystyle{
\times \, F_{\alpha}(P, {\bf s}, {\bf t}, \vec Z)\, \exp 
\Bigl (-\frac{1}{2}\sum_{i,j} \Omega_{ij}^{(\gamma \gamma )}\,
\frac{k_{\gamma}^{-i-j}}{ij}\Bigr ).
}
\end{array}
\eeq
The point $P$ in the r.h.s. of this
expression is assumed to belong to the neighborhood
of the point $P_{\gamma}$ parametrized by the local parameter $1/k_{\gamma}$,
and in this way the r.h.s. is regarded as a function of $k_{\gamma}$. 
At $\gamma =\alpha$ we have
\beq\label{stau4}
\begin{array}{l}
\displaystyle{
\psi_{\alpha \alpha}({\bf s}, {\bf t}, k_{\alpha})=
\exp \left (
\hat \Omega_{\alpha}(P, {\bf t})+
\sum_{\mu \neq \alpha}\Omega_0^{(\alpha \mu )}(P)s_{\mu}-
\sum_{\mu \neq \alpha}\Omega_{00}^{(\alpha \mu )}s_{\mu} \right )}
\\ \\
\phantom{aaaaaaaaaaaaaaaaaaaaaaaa}\displaystyle{
\times \, F_{\alpha}(P, {\bf s}, {\bf t}, \vec Z)\, \exp 
\Bigl (-\frac{1}{2}\sum_{i,j} \Omega_{ij}^{(\alpha \alpha )}\,
\frac{k_{\alpha}^{-i-j}}{ij}\Bigr ).
}
\end{array}
\eeq
Again, the point $P$ here
is assumed to belong to the neighborhood
of the point $P_{\alpha}$ parametrized by the local parameter $1/k_{\alpha}$,
and in this way the r.h.s. is regarded as a function of $k_{\alpha}$. 
A similar calculation for $\psi^*_{\gamma \beta}$ at $\gamma \neq \beta$
yields
\beq\label{stau5}
\begin{array}{l}
\displaystyle{
\psi^*_{\gamma \beta}({\bf s}', {\bf t}', k_{\gamma})=
\epsilon^{-1}_{\beta \gamma}({\bf s}')
\exp \left (-\frac{i\pi}{2}(s_{\gamma}'\! -\! s_{\beta}'\! +\! 1)-
\hat \Omega_{\beta}(P, {\bf t}')-
\sum_{\mu \neq \gamma}\Omega_0^{(\gamma \mu )}(P)s_{\mu}'+
\Omega_0^{(\gamma \beta )}(P) \right.}
\\ \\
\phantom{aaaaaaa}\displaystyle{
\left. +\frac{1}{2}\sum_{\mu \neq \beta}(\Omega_{00}^{(\mu \beta )}
+\Omega_{00}^{(\beta \mu )})s_{\mu}'-
\frac{1}{2}\sum_{\mu \neq \gamma}(\Omega_{00}^{(\mu \gamma )}
-\Omega_{00}^{(\gamma \mu )})s_{\mu}'+\frac{1}{2}(\Omega_{00}^{(\beta \gamma )}
-\Omega_{00}^{(\gamma \beta )})\right )}
\\ \\
\phantom{aaaaaaaaaaaaaaaaaaaaaaaa}\displaystyle{
\times \, F_{\alpha}(P, -{\bf s}', -{\bf t}', -\vec Z)\, \exp 
\Bigl (-\frac{1}{2}\sum_{i,j} \Omega_{ij}^{(\gamma \gamma )}\,
\frac{k_{\gamma}^{-i-j}}{ij}\Bigr ).
}
\end{array}
\eeq
At $\gamma =\beta$ we have
\beq\label{stau6}
\begin{array}{l}
\displaystyle{
\psi_{\beta \beta}({\bf s}', {\bf t}', k_{\alpha})=
\exp \left (
-\hat \Omega_{\beta}(P, {\bf t}')-
\sum_{\mu \neq \beta}\Omega_0^{(\beta \mu )}(P)s_{\mu}'+
\sum_{\mu \neq \beta}\Omega_{00}^{(\beta \mu )}s_{\mu}' \right )}
\\ \\
\phantom{aaaaaaaaaaaaaaaaaaaaaaaa}\displaystyle{
\times \, F_{\beta}(P, -{\bf s}', -{\bf t}', -\vec Z)\, \exp 
\Bigl (-\frac{1}{2}\sum_{i,j} \Omega_{ij}^{(\beta \beta )}\,
\frac{k_{\beta}^{-i-j}}{ij}\Bigr ).
}
\end{array}
\eeq
Therefore, at $\gamma \neq \alpha , \beta$ we obtain:
\beq\label{stau8}
\begin{array}{l}
\displaystyle{
\psi_{\alpha \gamma}({\bf s}, {\bf t}, k_{\gamma})
\psi_{\gamma \beta}^* ({\bf s}', {\bf t}', k_{\gamma})dk_{\gamma}}
\\ \\
\phantom{aaaaaaaaa}\displaystyle{=\epsilon_{\alpha \gamma}({\bf s})
\epsilon^{-1}_{\beta \gamma}({\bf s}')
\exp \left (\vphantom{\sum_{\mu \neq \alpha}^a}
-\frac{i\pi}{2}(s_{\alpha}\! -\! s_{\gamma}\! +\!
s_{\gamma}' \! -\! s_{\beta}'\! +\! 2)+\hat \Omega_{\alpha}(P, {\bf t})-
\hat \Omega_{\beta}(P, {\bf t}')
\right.
}
\\ \\
\phantom{aaaaaaaaaaaaaaaaa}
\displaystyle{+\sum_{\mu \neq \gamma}\Omega_0^{(\gamma \mu )}(P)(s_{\mu}-
s_{\mu}') -\Omega_0^{(\alpha \gamma )}(P)+\Omega_0^{(\gamma \beta )}(P)
}
\\ \\
\phantom{aaaaaaaaaaaaaaaaaaaaaaaaa}\displaystyle{
-\frac{1}{2}\sum_{\mu \neq \alpha}(\Omega_{00}^{(\mu \alpha )}+
\Omega_{00}^{(\alpha \mu )})s_{\mu} +
\frac{1}{2}\sum_{\mu \neq \beta}(\Omega_{00}^{(\mu \beta )}+
\Omega_{00}^{(\beta \mu )})s_{\mu}'
}
\\ \\
\phantom{aaaaaaaaaaaaaaa}\displaystyle{\left.
+\frac{1}{2}\sum_{\mu \neq \gamma}(\Omega_{00}^{(\mu \gamma )}-
\Omega_{00}^{(\gamma \mu )})(s_{\mu}-s'_{\mu})+
\frac{1}{2}\Bigl (\Omega_{00}^{(\alpha \gamma )}\! -\!
\Omega_{00}^{(\gamma \alpha )}\! +\!
\Omega_{00}^{(\beta \gamma )}\! -\!
\Omega_{00}^{(\gamma \beta )}\Bigr )\vphantom{\sum_{\mu \neq \alpha}^a}
\right )}
\\ \\
\phantom{aaaaaaaaaaaaaaa}\displaystyle{
\times \, F_{\alpha}(P, {\bf s}, {\bf t}, \vec Z)
F_{\beta}(P, -{\bf s}', -{\bf t}', -\vec Z)
\exp 
\Bigl (-\sum_{i,j} \Omega_{ij}^{(\gamma \gamma )}\,
\frac{k_{\gamma}^{-i-j}}{ij}\Bigr )dk_{\gamma}.
}
\end{array}
\eeq
For the last exponential factor in this expression we have
(see (\ref{stau7a})):
\beq\label{stau7}
\exp 
\Bigl (-\sum_{i,j} \Omega_{ij}^{(\gamma \gamma )}\,
\frac{k_{\gamma}^{-i-j}}{ij}\Bigr )dk_{\gamma}=
\frac{C_{\gamma}\, d\zeta (P)}{\Theta_{*}^2(\vec A(P)-\vec A(P_{\gamma}))},
\eeq
where $d\zeta$ is the holomorphic differential (\ref{qp86}). 

The comparison with $\Psi_{\alpha}({\bf s}, {\bf t}, P)
\Psi^*_{\beta}({\bf s}', {\bf t}', P) d\Omega$ gives:
$$
\psi_{\alpha \gamma}({\bf s}, {\bf t}, k_{\gamma})
\psi_{\gamma \beta}^* ({\bf s}', {\bf t}', k_{\gamma})dk_{\gamma}
\phantom{aaaaaaaaaaaaaaa}
$$
$$
=
C_{\alpha \beta}^{-1} A_{\alpha}({\bf s})A_{\beta}^{-1}({\bf s}')
\Psi_{\alpha}({\bf s}, {\bf t}, P)
\Psi^*_{\beta}({\bf s}', {\bf t}', P) d\Omega
$$
$$
\times \, \epsilon_{\alpha \gamma}({\bf s})
\epsilon^{-1}_{\beta \gamma}({\bf s}')
\exp \left (\frac{i\pi}{2}( \! s_{\gamma}\! -\!
s_{\gamma}' \!  -\! 2)
+\frac{1}{2}\sum_{\mu \neq \gamma}(L_{\mu \gamma}-L_{\gamma \mu })s_{\mu}-
\frac{1}{2}\sum_{\mu \neq \gamma}(L_{\mu \gamma}-L_{\gamma \mu })s_{\mu}'
\right.
$$
$$
+\,\left. \vphantom{\frac{1}{2}\sum_{\mu \neq \gamma}}
\frac{1}{2} \Bigl (L_{\alpha \gamma}\! -\! L_{\gamma \alpha}\! +\!
L_{\beta \gamma}\! -\! L_{\gamma \beta}\Bigr )\right ),
$$
where
\beq\label{stau9}
\begin{array}{l}
\displaystyle{
A_{\alpha}({\bf s})=\exp \left (-\frac{i\pi}{2}s_{\alpha}+
\sum_{\mu \neq \alpha}\Omega_0^{(0\mu )}
(P_{\alpha}) s_{\mu}-\Omega_{00}^{(\alpha 0)} s_{\alpha}\right. }
\\ \\
\phantom{aaaaaaaaaaaaaaaaaaa}\displaystyle{ \left.
-\frac{1}{2}\sum_{\mu \neq \alpha}(\Omega_{00}^{(\mu \alpha )}+
\Omega_{00}^{(\alpha \mu )})s_{\mu}+\frac{1}{2}\sum_{\mu}\log C_{\mu}\,
s_{\mu}\right ),}
\end{array}
\eeq

\beq\label{stau10}
L_{\alpha \beta}=\log \Theta_{*}(\vec A(P_{\alpha})-\vec A(P_{\beta})).
\eeq

\noindent
In a similar way, we obtain the following identities for the cases
when $\gamma =\alpha$, $\gamma =\beta$ and $\gamma =\alpha =\beta$:
$$
\begin{array}{c}
\displaystyle{
\psi_{\alpha \alpha}({\bf s}, {\bf t}, k_{\alpha})
\psi_{\alpha \beta}^* ({\bf s}', {\bf t}', k_{\alpha})dk_{\alpha}=
C_{\alpha \beta}^{-1} A_{\alpha}({\bf s})A_{\beta}^{-1}({\bf s}')
\Psi_{\alpha}({\bf s}, {\bf t}, P)
\Psi^*_{\beta}({\bf s}', {\bf t}', P) d\Omega }
\\ \\
\displaystyle{
\times \, \epsilon^{-1}_{\beta \alpha}({\bf s}')
\exp \left (\frac{i\pi}{2}(s_{\alpha}\! -\! s_{\alpha}'\! -\! 1)
+\frac{1}{2}\sum_{\mu \neq \alpha}(L_{\mu \alpha}-L_{\alpha \mu })(s_{\mu}-
s_{\mu}')
+\frac{1}{2}(L_{\beta \alpha}\! -\! L_{\alpha \beta})
\right ),}
\end{array}
$$

$$
\begin{array}{l}
\displaystyle{
\psi_{\alpha \beta}({\bf s}, {\bf t}, k_{\alpha})
\psi_{\beta \beta}^* ({\bf s}', {\bf t}', k_{\beta})dk_{\beta}=
C_{\alpha \beta}^{-1} A_{\alpha}({\bf s})A_{\beta}^{-1}({\bf s}')
\Psi_{\alpha}({\bf s}, {\bf t}, P)
\Psi^*_{\beta}({\bf s}', {\bf t}', P) d\Omega }
\\ \\
\phantom{aaaaa}\displaystyle{
\times \, \epsilon_{\alpha \beta}({\bf s}) \exp \left (\frac{i\pi}{2}(
s_{\beta} \! -\! s_{\beta}'\! -\! 1)
+\frac{1}{2}\sum_{\mu \neq \beta}(L_{\mu \beta}-L_{\beta \mu })(s_{\mu}-
s_{\mu}')
+\frac{1}{2}(L_{\alpha \beta}\! -\! L_{\beta \alpha})
\right ),}
\end{array}
$$

$$
\psi_{\alpha \alpha}({\bf s}, {\bf t}, k_{\alpha})
\psi_{\alpha \alpha}^* ({\bf s}', {\bf t}', k_{\alpha})dk_{\alpha}=
C_{\alpha \alpha}^{-1} A_{\alpha}({\bf s})A_{\alpha}^{-1}({\bf s}')
\Psi_{\alpha}({\bf s}, {\bf t}, P)
\Psi^*_{\alpha}({\bf s}', {\bf t}', P) d\Omega
$$
$$
\phantom{aaaaaaaaaaa}
\times \, \exp \left (\frac{i\pi}{2}(s_{\alpha}\! -\! s_{\alpha}')+
\frac{1}{2}\sum_{\mu \neq \alpha}(L_{\mu \alpha}-L_{\alpha \mu })(s_{\mu}-
s_{\mu}')
\right ).
$$
Let us consider the factors containing the $L$-functions (\ref{stau10})
and depending on ${\bf s}$, ${\bf s}'$.
To find them
explicitly, we need to specify the branches of the logarithmic functions
(\ref{stau10}). It is clear that we can not put $L_{\alpha \beta}-
L_{\beta \alpha}=i\pi$ for all $\alpha \neq \beta$ because
if $L_{12}-L_{21}=i\pi$, then $L_{21}-L_{12}=-i\pi$. This suggests to
fix the branches as follows:
\beq\label{stau11}
L_{\alpha \beta}-L_{\beta \alpha}=\left \{
\begin{array}{rl} i\pi & \mbox{if $\alpha <\beta$,}
\\
-i\pi & \mbox{if $\alpha >\beta$.}
\end{array}\right.
\eeq
Vice versa, any choice of the branches fixes an order 
in the set of the indices $1, \ldots , n$. 
Then a simple calculation which uses (\ref{s1}) yields
\beq\label{stau12}
\frac{i\pi}{2}s_{\alpha}+\frac{1}{2}\sum_{\mu \neq \alpha}(L_{\mu \alpha}-
L_{\alpha \mu})s_{\mu}=-i\pi\sum_{\mu >\alpha} s_{\mu}.
\eeq
Besides, we have
\beq\label{stau13}
-\frac{i\pi}{2} +\frac{1}{2}\, (L_{\alpha \beta} -L_{\beta \alpha})=
\left \{ \begin{array}{cl}
0 & \mbox{if $\alpha <\beta$,}
\\ 
-i\pi & \mbox{if $\alpha >\beta$.}
\end{array}\right.
\eeq
Therefore, we see that
\beq\label{stau14}
\exp \left (\frac{i\pi}{2}s_{\gamma}+\frac{1}{2}
\sum_{\mu \neq \gamma}(L_{\mu \gamma}-
L_{\gamma \mu})s_{\mu}-\frac{i\pi}{2} +\frac{1}{2}\, 
(L_{\alpha \gamma} -L_{\gamma \alpha})\right )=\epsilon^{-1}_{\alpha \gamma}
({\bf s})\exp \left (i\pi \sum_{\mu \leq \alpha}s_{\mu}\right ),
\eeq
where $\epsilon_{\alpha \gamma}
({\bf s})$ is given by (\ref{s3a}), and so
\beq\label{stau15}
\begin{array}{l}
\psi_{\alpha \gamma}({\bf s}, {\bf t}, k_{\gamma})
\psi_{\gamma \beta}^* ({\bf s}', {\bf t}', k_{\gamma})dk_{\gamma}
\\ \\
\phantom{aaaaaaaaaaaaa}=
C_{\alpha \beta}^{-1} B_{\alpha}({\bf s})B_{\beta}^{-1}({\bf s}')
\Psi_{\alpha}({\bf s}, {\bf t}, P)
\Psi^*_{\beta}({\bf s}', {\bf t}', P) d\Omega
\end{array}
\eeq
for all $\alpha$, $\beta$, $\gamma$. Here
\beq\label{stau16}
B_{\alpha}({\bf s})=\exp \left (i\pi \sum_{\mu\leq \alpha}s_{\mu}\right )
A_{\alpha}({\bf s}).
\eeq
Note that the r.h.s. of (\ref{stau15}) depends
on $\gamma$ only through the point $P$ which belongs to a neighborhood
of the point $P_{\gamma}$ with the local parameter $k_{\gamma}^{-1}$. 
Taking the sum of residues of the r.h.s. at the points $P_{\gamma}$ for
$\gamma =1, \ldots , n$ (which is $0$ according to (\ref{v10})), we get
the bilinear equation (\ref{s5}). We have proved that the tau-function
(\ref{stau1}) is indeed a solution to the bilinear equation (\ref{s3}).

\section{Multi-component KP and Toda hierarchies}

\subsection{Multi-component KP hierarchy}

The multi-component KP hierarchy is obtained from the universal
hierarchy by restricting the ${\bf s}$-variables to zero values.
Accordingly, when discussing the multi-component 
KP hierarchy we skip ${\bf s}$ in the notation, writing simply
$\tau ({\bf t})$, etc.

The tau-function of the multi-component KP hierarchy 
can be regarded as an $n\! \times \! n$ matrix
with matrix elements
\beq\label{m1}
\tau_{\alpha \beta}({\bf t})=\left \{
\begin{array}{ll}
\tau ([1]_{\alpha \beta}, {\bf t}) & \mbox{if $\alpha \neq \beta$},
\\ & \\
\tau (0, {\bf t}) & \mbox{if $\alpha = \beta$}.
\end{array} \right.
\eeq
The $n$-component KP hierarchy is the infinite set of bilinear equations
for the tau-functions which are encoded in the basic bilinear
relation
\beq\label{m5}
\sum_{\gamma =1}^n \epsilon_{\alpha \gamma}\epsilon_{\beta \gamma}
\oint_{C_{\infty}}\! dk \, 
k^{\delta_{\alpha \gamma}+\delta_{\beta \gamma}-2}
e^{\xi ({\bf t}_{\gamma}-{\bf t}_{\gamma}', \, k)}
\tau _{\alpha \gamma} \left ({\bf t}-[k^{-1}]_{\gamma}\right )
\tau _{\gamma \beta}\left ({\bf t}'+[k^{-1}]_{\gamma}\right )=0
\eeq 
valid for all ${\bf t}$, ${\bf t}'$. Here $\epsilon_{\alpha \gamma}=1$
if $\alpha \leq \gamma$ and $\epsilon_{\alpha \gamma}=-1$ 
if $\alpha >\gamma$.

The algebraic-geometrical tau-function $\tau_{\alpha \beta}({\bf t})$
which solves (\ref{m5})
is obtained from the universal tau-function given by (\ref{stau1}),
(\ref{stau2}) by restriction to the ${\bf s}=0$ sector. 
We have:
\beq\label{vb15}
\tau_{\alpha \beta}({\bf t})=\left (
\frac{C_{\alpha}\epsilon_{\alpha \beta}}{\Theta_*(\vec A(P_{\alpha})\! -\! 
\vec A(P_{\beta}))}\right )^{1-\delta_{\alpha \beta}}\!\!
e^{-Q_{\alpha \beta}({\bf t})}\, \Theta (\vec A(P_{\beta})-
\vec A(P_{\alpha})+\vec U({\bf t})+\vec Z),
\eeq
where
\beq\label{mtau2}
Q_{\alpha \beta}({\bf t})=\frac{1}{2}\sum_{\mu , \nu}\sum_{i,j}
\Omega_{ij}^{(\mu \nu )}t_{\mu , i}t_{\nu , j}+
\sum_{\mu \neq \alpha}\sum_j \Omega_{0j}^{(\mu \alpha )}t_{\mu , j}-
\sum_{\mu \neq \beta}\sum_j \Omega_{0j}^{(\mu \beta )}t_{\mu , j}.
\eeq

\subsection{Multi-component Toda hierarchy: bilinear formalism}

Let us show that $n=2N$-component universal hierarchy contains 
the $N$-component Toda hierarchy \cite{UT84,TZ}. 

The independent variables of the multi-component Toda hierarchy are $2N$ 
infinite sets of ``times''
$$
{\bf t}=\{{\bf t}_1, {\bf t}_2, \ldots , {\bf t}_N\}, \qquad
{\bf t}_{\alpha}=\{t_{\alpha , 1}, t_{\alpha , 2}, 
t_{\alpha , 3}, \ldots \, \},
\qquad \alpha = 1, \ldots , N
$$
and
$$
\bar {\bf t}=\{\bar {\bf t}_1, \bar {\bf t}_2, \ldots , 
\bar {\bf t}_N\}, \qquad
\bar {\bf t}_{\alpha}=\{ \bar t_{\alpha , 1}, \bar t_{\alpha , 2}, 
\bar t_{\alpha , 3}, \ldots \, \},
\qquad \alpha = 1, \ldots , N
$$
which are in general complex numbers.
Besides, there are variables $n_1, \ldots , n_N$ which are
usually regarded as integers but in algebraic-geometrical 
solutions can be treated as complex numbers. This set
is denoted as ${\bf n}=\{n_1, \ldots , n_N\}$. 

Let the set of $2N$ indices in the universal hierarchy be
$\{1, 2, \ldots , N, \bar 1, \bar 2, \ldots , \bar N\}$. We identify
the time variables ${\bf t}_{\alpha}$ with ${\bf t}_{\alpha}$ of the 
universal hierarchy and
$\bar {\bf t}_{\alpha}$ with ${\bf t}_{\bar \alpha}$ 
for $\alpha =1, \ldots , N$. Besides, we set 
$s_{\alpha}=-s_{\bar \alpha}=n_{\alpha}$. By ${\bf T}$ we denote
the set of times ${\bf T}=\{{\bf t}_1, {\bf t}_2, \ldots , {\bf t}_N,
\bar {\bf t}_1, \bar {\bf t}_2, \ldots , \bar {\bf t}_N\}$ and by
$\bar {\bf n}$ we denote
the $2N$-component set $\bar {\bf n}=\{n_1, n_2, \ldots , n_N, -n_1, -n_2,
\ldots , -n_N\}$. Consider the tau-function 
$\tau (\bar {\bf n}+{\bf r}, {\bf T})$ of the $2N$-component universal
hierarchy, where the $2N$ 
variables ${\bf r}$ are subject to the constraint
$$
\sum_{\nu =1}^N (r_{\nu}+r_{\bar \nu})=0.
$$ 
The tau-function of the $N$-component Toda hierarchy
is the $N\times N$ matrix with matrix elements
\beq\label{mt1}
\tau_{\alpha \gamma} ({\bf n}, {\bf t}, \bar {\bf t})=
\tau (\bar {\bf n}+[1]_{\alpha \gamma}, {\bf T}), \quad \alpha , \gamma =
1, \ldots , N.
\eeq
It is easy to see that
\beq\label{mt2}
\begin{array}{l}
\tau (\bar {\bf n}+[1]_{\alpha \bar \gamma}, {\bf T})=
\tau_{\alpha \gamma} ({\bf n}+[1]_{\gamma}, {\bf t}, \bar {\bf t}),
\\ \\
\tau (\bar {\bf n}+[1]_{\bar \alpha \gamma}, {\bf T})=
\tau_{\alpha \gamma} ({\bf n}-[1]_{\alpha}, {\bf t}, \bar {\bf t}),
\\ \\
\tau (\bar {\bf n}+[1]_{\bar \alpha \bar \gamma}, {\bf T})=
\tau_{\alpha \gamma} ({\bf n}+[1]_{\gamma \alpha}, {\bf t}, \bar {\bf t}),
\end{array}
\eeq
where
$$
{\bf n}\pm [1]_{\gamma}=\{n_1, \ldots , n_{\gamma -1}, n_{\gamma}\pm 1,
n_{\gamma +1}, \ldots , n_N\},
\quad 
{\bf n}+ [1]_{\gamma \alpha}={\bf n}+ [1]_{\gamma}-[1]_{\alpha}.
$$

It is not difficult to verify that in this setting 
the basic bilinear equation (\ref{s3}) acquires the following form:
\beq\label{mt3b}
\begin{array}{l}
\displaystyle{
\sum_{\gamma =1}^N 
\epsilon_{\beta \gamma}
\epsilon_{\alpha \gamma}({\bf n})
\epsilon^{-1}_{\beta \gamma}({\bf n}')
\oint_{C_{\infty}}
\! dk \, k^{n_{\gamma}-n_{\gamma}' +\delta_{\alpha \gamma}-2}
e^{\xi ({\bf t}_{\gamma}-{\bf t}_{\gamma}', k)}}
\\ \\
\phantom{aaaaaaaaaaaaaaaa}
\displaystyle{
\times \tau_{\alpha \gamma} ({\bf n}, 
{\bf t}-[k^{-1}]_{\gamma}, \bar {\bf t})
\tau_{\gamma \beta}({\bf n}'+[1]_{\beta}, 
{\bf t}'+[k^{-1}]_{\gamma}, \bar {\bf t}')}
\\ \\
\displaystyle{
=\, \sum_{\gamma =1}^N
\epsilon_{\alpha \gamma}
\epsilon_{\alpha \gamma}({\bf n})
\epsilon^{-1}_{0N}({\bf n})
\epsilon^{-1}_{\beta \gamma}({\bf n}')
\epsilon_{0N}({\bf n}')
\oint_{C_{\infty}}
\! dk \, k^{n_{\gamma}'-n_{\gamma} +\delta_{\beta \gamma}-2}
e^{\xi (\bar {\bf t}_{\gamma}-\bar {\bf t}_{\gamma}', k)}}
\\ \\
\phantom{aaaaaaaaaaaa}
\displaystyle{
\times \tau_{\alpha \gamma} ({\bf n}+[1]_{\gamma}, 
{\bf t}, \bar {\bf t}-[k^{-1}]_{\gamma})
\tau_{\gamma \beta}({\bf n}'+[1]_{\beta \gamma}, 
{\bf t}', \bar {\bf t}'+[k^{-1}]_{\gamma})},
\end{array}
\eeq
where
$$
\epsilon_{0N}({\bf n})=\exp \Bigl (-i\pi
\sum_{1\leq \mu \leq N}n_{\mu}\Bigr ).
$$
This is the bilinear equation for the tau-function of the 
$N$-component Toda hierarchy.
It holds for all ${\bf t}$, ${\bf t}'$, $\bar {\bf t}$,
$\bar {\bf t}'$, ${\bf n}$, ${\bf n}'$ such that 
${\bf n}-{\bf n}'\in \ZZ^N$.

For $N=1$ equation (\ref{mt3b}) becomes the standard bilinear
equation
\beq\label{toda}
\begin{array}{l}
\displaystyle{
\oint_{C_{\infty}}
\! dk \, k^{n-n'-1}
e^{\xi ({\bf t}-{\bf t}', k)}
\tau^{\rm Toda}_{n} ( 
{\bf t}-[k^{-1}], \bar {\bf t})
\tau^{\rm Toda}_{n'+1}( 
{\bf t}'+[k^{-1}], \bar {\bf t}')}
\\ \\
\displaystyle{
=\oint_{C_{\infty}}
\! dk \, k^{n'-n-1}
e^{\xi (\bar {\bf t}-\bar {\bf t}', k)}
\tau^{\rm Toda}_{n+1} ( 
{\bf t}, \bar {\bf t}-[k^{-1}])
\tau^{\rm Toda}_{n'}( 
{\bf t}', \bar {\bf t}'+[k^{-1}])}
\end{array}
\eeq
for the tau-function $\tau^{\rm Toda}_{n} ( 
{\bf t}, \bar {\bf t})=e^{\frac{\pi i}{2}\, n^2}\tau_{11}(n, {\bf t}, 
\bar {\bf t})$ of the Toda lattice hierarchy \cite{UT84}. 

In terms of the $\psi$-functions
\beq\label{e2}
\begin{array}{l}
\displaystyle{
\psi_{\alpha \gamma}({\bf n}, {\bf t}, \bar {\bf t};k)=
\epsilon_{\alpha \gamma}({\bf n}) k^{n_{\gamma}+\delta_{\alpha \gamma}-1}
e^{\xi ({\bf t}_{\gamma}, k)}\frac{\tau_{\alpha \gamma}
({\bf n}, {\bf t}-[k^{-1}]_{\gamma}, \bar {\bf t})}{\tau
({\bf n}, {\bf t}, \bar {\bf t})},}
\\ \\
\displaystyle{
\psi^*_{\gamma \beta}({\bf n}, {\bf t}, \bar {\bf t};k)=
\epsilon_{\beta \gamma}
\epsilon^{-1}_{\beta \gamma}({\bf n}) 
k^{-n_{\gamma}}
e^{-\xi ({\bf t}_{\gamma}, k)}\frac{\tau_{\gamma \beta}
({\bf n}+[1]_{\beta}, {\bf t}+[k^{-1}]_{\gamma}, \bar {\bf t})}{\tau
({\bf n}, {\bf t}, \bar {\bf t})},}
\\ \\
\displaystyle{
\bar \psi_{\alpha \gamma}({\bf n}, {\bf t}, \bar {\bf t};k)=
\epsilon_{0N}^{-1}({\bf n})\epsilon_{\alpha \gamma}
\epsilon_{\alpha \gamma}({\bf n}) k^{-n_{\gamma}}
e^{\xi (\bar {\bf t}_{\gamma}, k)}\frac{\tau_{\alpha \gamma}
({\bf n}+[1]_{\gamma}, {\bf t}, \bar {\bf t}-[k^{-1}]_{\gamma})}{\tau
({\bf n}, {\bf t}, \bar {\bf t})},}
\\ \\
\displaystyle{
\bar \psi^*_{\gamma \beta}({\bf n}, {\bf t}, \bar {\bf t};k)=
\epsilon_{0N}({\bf n})
\epsilon^{-1}_{\beta \gamma}({\bf n}) 
k^{n_{\gamma}+\delta_{\beta \gamma}-1}
e^{-\xi (\bar {\bf t}_{\gamma}, k)}\frac{\tau_{\gamma \beta}
({\bf n}+[1]_{\beta \gamma}, {\bf t}, \bar {\bf t}+
[k^{-1}]_{\gamma})}{\tau
({\bf n}, {\bf t}, \bar {\bf t})}}
\end{array}
\eeq
the bilinear relation (\ref{mt3b}) has the form
\beq\label{e1}
\sum_{\gamma =1}^N \oint_{C_{\infty}}\!\! \frac{dk}{k}
\psi_{\alpha \gamma}({\bf n}, {\bf t}, \bar {\bf t};k)
\psi^*_{\gamma \beta} ({\bf n}', {\bf t}', \bar {\bf t}';k)
=
\sum_{\gamma =1}^N \oint_{C_{\infty}}\!\! \frac{dk}{k}
\bar \psi_{\alpha \gamma}({\bf n}, {\bf t}, \bar {\bf t};k)
\bar \psi^*_{\gamma \beta} ({\bf n}', {\bf t}', \bar {\bf t}';k).
\eeq

\subsection{Multi-component Toda hierarchy: 
quasi-periodic solutions}

The vector Baker-Akhiezer functions for the non-abelian Toda
lattice were introduced in \cite{KZ95}. Here we generalize them,
with some modifications, to the $N$-component Toda hierarchy.

Let us fix $2N$ marked points $P_1, \ldots , P_N$ and
$\bar P_1, \ldots , \bar P_N$ 
on the curve $\Gamma$ with local parameters $k^{-1}_{\gamma},
\bar k^{-1}_{\gamma}$, $\gamma =1, \ldots , N$ (the bar here does not mean
complex conjugation),
and make $N$ (non-intersecting) 
cuts ${\sf c}_{\alpha}$ from $P_{\alpha}$ to $\bar P_{\alpha}$.
We also fix $g+N-1$ points $Q_1 , \ldots , Q_{g+N-1}$ 
in general position. The corresponding effective divisor is
$$
{\cal D}=Q_1+Q_2+\ldots +Q_{g+N-1}.
$$

\begin{proposition}
There exists a unique
function $\Phi_{\alpha}({\bf n},{\bf t}, \bar {\bf t}, P)$ 
on $\Gamma$ such that:
\begin{itemize}
\item[$1^{0}$.] The function 
$\Phi_{\alpha}({\bf n},{\bf t}, \bar {\bf t}, P)$ is meromorphic outside the
marked points 
and cuts and has at most simple poles at the points of the divisor
${\cal D}$ (if all of them are distinct);
\item[$2^{0}$.] The boundary values 
$\Phi^{(\pm )}_{\alpha}({\bf n},{\bf t}, \bar {\bf t}, P)$ 
of the function $\Phi_{\alpha}$ 
on the different sides 
of the cut ${\sf c}_{\beta}$ satisfy the relation
\beq\label{v0b}
\Phi_{\alpha}^{(+ )}
({\bf n},{\bf t}, \bar {\bf t}, P)=e^{2\pi i n_{\beta}}
\Phi_{\alpha}^{(-)}
({\bf n},{\bf t}, \bar {\bf t}, P), \qquad P\in {\sf c}_{\beta};
\eeq
\item[$3^{0}$.] In the neighborhood of the marked points
$P_{\beta}$, $\bar P_{\beta}$ ($k_{\beta}, \bar k_{\beta}\to \infty$)
it has the 
form
\beq\label{v1b}
\begin{array}{l}
\displaystyle{
\Phi_{\alpha}({\bf n},{\bf t}, \bar {\bf t}, P)=k_{\beta}^{n_{\beta}}
e^{\xi ({\bf t}_{\beta}, k_{\beta})}
\left (\delta_{\alpha \beta} +\sum_{j\geq 1}\xi_{\alpha \beta , j}
({\bf n}, {\bf t}, \bar {\bf t})k_{\beta}^{-j}\right ),}
\\ \\
\displaystyle{
\Phi_{\alpha}({\bf n},{\bf t}, \bar {\bf t}, P)=\bar k_{\beta}^{-n_{\beta}}
e^{\xi (\bar {\bf t}_{\beta}, \bar k_{\beta})}
\left (\bar \xi_{\alpha \beta , 0}
({\bf n}, {\bf t}, \bar {\bf t}) +\sum_{j\geq 1}\bar 
\xi_{\alpha \beta , j}
({\bf n}, {\bf t}, \bar {\bf t})\bar k_{\beta}^{-j}\right ).}
\end{array}
\eeq
\end{itemize}
\end{proposition}

\noindent
Comparing to the vector Baker-Akhiezer function $\Psi_{\alpha}$ for the
multi-component KP hierarchy with the same set of marked points, 
which has
$2N$ simple poles, the function $\Phi_{\alpha}$ has only $N$ simple poles
and is normalized in a different way.

The definition of the dual Baker-Akhiezer function is also different.
Let $${\cal D}^{*}=Q_1^{*}+ \ldots + Q_{g+N-1}^{*}$$ be the effective 
divisor satisfying the condition
\beq\label{v2b}
{\cal D}+{\cal D}^{*}={\cal K}+\sum_{\gamma =1}^N P_{\gamma}+
\sum_{\gamma =1}^N \bar P_{\gamma},
\eeq
where ${\cal K}$ is the canonical class. 
In terms of the Abel map, this condition reads
\beq\label{v2c}
\vec A({\cal D})+\vec A({\cal D}^{*})+\vec K=\sum_{\gamma =1}^N
\vec A(P_{\gamma})
+\sum_{\gamma =1}^N
\vec A(\bar P_{\gamma}),
\eeq
where $\vec K$ is the vector of Riemann's constants.

\begin{proposition}
There exists a unique
function $\Phi^*_{\alpha}({\bf n},{\bf t}, \bar {\bf t}, P)$ 
on $\Gamma$ such that:
\begin{itemize}
\item[$1^{0}$.] The function 
$\Phi^*_{\alpha}({\bf n},{\bf t}, \bar {\bf t}, P)$ 
is meromorphic outside the
marked points 
and cuts and has at most simple poles at the points of the divisor
${\cal D}^*$ (if all of them are distinct);
\item[$2^{0}$.] The boundary values 
$\Phi^{*(\pm )}_{\alpha}({\bf n},{\bf t}, \bar {\bf t}, P)$ 
of the function $\Phi^*_{\alpha}$ 
on the different sides 
of the cut ${\sf c}_{\beta}$ satisfy the relation
\beq\label{v0d}
\Phi_{\alpha}^{*(+ )}
({\bf n},{\bf t}, \bar {\bf t}, P)=e^{-2\pi i n_{\beta}}
\Phi_{\alpha}^{*(-)}
({\bf n},{\bf t}, \bar {\bf t}, P), \qquad P\in {\sf c}_{\beta};
\eeq
\item[$3^{0}$.] In the neighborhood of the marked points
$P_{\beta}$, $\bar P_{\beta}$ ($k_{\beta}, \bar k_{\beta}\to \infty$)
it has the 
form
\beq\label{v1e}
\begin{array}{l}
\displaystyle{
\Phi^*_{\alpha}({\bf n},{\bf t}, \bar {\bf t}, P)=k_{\beta}^{-n_{\beta}}
e^{-\xi ({\bf t}_{\beta}, k_{\beta})}
\left (\xi^*_{\alpha \beta , 0}
({\bf n}, {\bf t}, \bar {\bf t}) +\sum_{j\geq 1}\xi^*_{\alpha \beta , j}
({\bf n}, {\bf t}, \bar {\bf t})k_{\beta}^{-j}\right ),}
\\ \\
\displaystyle{
\Phi^*_{\alpha}({\bf n},{\bf t}, \bar {\bf t}, P)=\bar k_{\beta}^{n_{\beta}}
e^{-\xi (\bar {\bf t}_{\beta}, \bar k_{\beta})}
\left (\delta_{\alpha \beta}+
\sum_{j\geq 1}
\bar \xi^*_{\alpha \beta , j}
({\bf n}, {\bf t}, \bar {\bf t})\bar k_{\beta}^{-j}\right ).}
\end{array}
\eeq
\end{itemize}
\end{proposition}

Let $d\Omega (P)$ be the meromorphic differential with simple poles
at the marked points $P_{\alpha}$, $\bar P_{\alpha}$ 
and zeros at the divisors ${\cal D}$, 
${\cal D}^*$. (For such differential we use the same notation as in section
4 but hope that this will not lead to a misunderstanding.)
Such a differential exists because of the condition (\ref{v2b}).
Moreover, it is unique up to multiplication by a constant factor. 
Indeed, the linear space of all differentials with simple poles
at the marked points has dimension $g+2N-1$ (there are 
$2N$ coefficients 
in front of $dk_{\alpha}/k_{\alpha}$, $d\bar k_{\alpha}/\bar k_{\alpha}$
$\alpha =1, \ldots , n$, with one condition that their sum, as the sum
of residues, is equal to zero, and $g$ coefficients
in front of the holomorphic differentials $d\omega_{\nu}$)
but fixing the positions of zeros 
imposes $g+2N-2$ conditions (the $2(g+N-1)$ zeros 
are subject to $g$ conditions (\ref{v2c})), 
so the space of such differentials with 
fixed zeros is one-dimensional. Alternatively, one may fix the divisor
${\cal D}$ and impose the condition that near each point 
$P_{\alpha}$ the differential $d\Omega$ behaves as 
$d\Omega (P)=dk_{\alpha}/k_{\alpha}+\ldots$. These conditions fix the dual divisor ${\cal D}^*$. 

For ${\bf n}-{\bf n}'\in \ZZ^N$ 
the differential $\Phi_{\alpha}({\bf n}, {\bf t}, \bar {\bf t}, P)
\Phi^*_{\beta}({\bf n}', {\bf t}', \bar {\bf t}', P)d\Omega (P)$
is meromorphic on $\Gamma$ 
outside the marked points, so the sum of its residues
is equal to zero:
\beq\label{res}
\begin{array}{l}
\displaystyle{
\sum_{\gamma =1}^N\res_{P=P_{\gamma}}\Bigl (\Phi _{\alpha}
({\bf n}, {\bf t}, \bar {\bf t}, P)
\Phi^*_{\beta}({\bf n}', {\bf t}', \bar {\bf t}', P)d\Omega (P)\Bigr )}
\\ \\
\displaystyle{\phantom{aaaaaaaaa}+\,
\sum_{\gamma =1}^N\res_{P=\bar P_{\gamma}}\Bigl (\Phi _{\alpha}
({\bf n}, {\bf t}, \bar {\bf t}, P)
\Phi^*_{\beta}({\bf n}', {\bf t}', \bar {\bf t}', P)d\Omega (P)\Bigr )=0}.
\end{array}
\eeq

The tau-function of the $N$-component Toda hierarchy 
for algebraic-geometrical solutions is the specialization
of the tau-function (\ref{stau1}) for ${\bf s}=\bar {\bf n}$. It is the
$N\times N$ 
matrix with matrix elements
\beq\label{tautoda}
\tau_{\alpha \beta} ({\bf n}, {\bf t}, \bar {\bf t})=
e^{-Q_{\alpha \beta }({\bf n}, {\bf t}, \bar {\bf t})}
\Theta \Bigl (\vec U({\bf t}, \bar {\bf t})+\vec U_0({\bf n})-
\vec A(P_{\alpha})+\vec A(\bar P_{\beta})+\vec Z \Bigr ),
\eeq
where
$$
\vec U({\bf t}, \bar {\bf t})=\sum_{\mu =1}^N \sum_{j\geq 1}
\vec U_j^{(\mu )}t_{\mu , j}+
\sum_{\mu =1}^N \sum_{j\geq 1}
\vec U_j^{(\bar \mu )}\bar t_{\mu , j},
$$
$$
\vec U_0({\bf n})=\sum_{\mu =1}^N n_{\mu}(\vec A(\bar P_{\mu})-
\vec A(P_{\mu}))=\sum_{\mu =1}^N n_{\mu} \vec U_0^{(\mu )}
$$
and $Q_{\alpha \beta }({\bf n}, {\bf t}, \bar {\bf t})$ is a quadratic
form in the times which is obtained as the specialization of (\ref{stau2}).
As a corollary of Theorem 5.1 we conclude that this tau-function satisfies
the bilinear equation (\ref{mt3b}), which is a direct consequence 
of (\ref{res}).

\section{Conclusions}

We have constructed quasi-periodic (algebraic-geometrical) solutions
to the universal integrable 
hierarchy of differential-difference equations which is the $n$-component
KP hierarchy extended by $n$ discrete flows. The universal
hierarchy contains both KP and Toda hierarchy as well as 
their mumti-component generalizations, and that is why we call it
universal. The quasi-periodic solutions are constructed starting from
a Riemann surface $\Gamma$ 
of finite genus $g$ with $n$ marked points together 
with local parameters at these points and an effective divisor 
of degree $g+n-1$. The tau-function is the Riemann theta-function 
of a linear combination of all the hierarchical times 
$t_{\alpha , k}$, with coefficients 
being periods of certain meromorphic differentials on $\Gamma$, multiplied
by exponential function of a quadratic form in the times. 

The main theorem proved in this paper (Theorem 5.1) states that the
tau-function mentioned above satisfies the integral bilinear equation
for the tau-function which was obtained in the works of Kyoto school.
The proof is based on a direct computation which uses certain 
non-trivial properties of differentials on Riemann surfaces. 
This result was expected but the proof, to the best of our knowledge, 
never appeared in the literature in an explicit form. 
The main technical tools in our proof are the vector 
Baker-Akhiezer function and its dual.

We have also specialized the result to the multi-component KP and Toda
hierarchies and present their matrix tau-functions in the explicit
form.

\section*{Acknowledgments}

\addcontentsline{toc}{section}{Acknowledgments}

The work of A.Z. is an output of a research project 
implemented as a part of the Basic Research Program 
at the National Research University Higher School 
of Economics (HSE University).

\end{document}